\input phyzzx.tex
\tolerance=1000
\voffset=-0.0cm
\hoffset=0.7cm
\sequentialequations
\def\rl{\rightline}

\def\t1{{\tilde 1}}

\def\t{\theta}

\def\S{Schwarzschild~}

\REF{\HOL}{G. 't Hooft, [arXiv:gr-qc/9310026]; L. Susskind, J. Math. Phys. {\bf 36} (1995) 6377, [arXiv:hep-th/9409089]; R. Bousso, Rev. Mod. Phys. {\bf 74} (2002) 825, [arXiv:hep-th/0203101].}
\REF{\ADS}{J. Maldacena, Adv. Theor. Math. Phys. {\bf 2} (1998) 231, [arXiv:hep-th/9711200]; S. Gubser, I. Klebanov and A. Polyakov, Phys. Lett. {\bf B428} (1998) 105, [arXiv:hep-th/9802109]; E. Witten, Adv. Theor. Math. Phys. {\bf 2} (1998) 253, [arXiv:hep-th/9802150].}
\REF{\EDI}{Edi Halyo, [arXiv:1706.07428].}
\REF{\WALD}{R. M. Wald, General Relativity, University of Chicago Press.}
\REF{\GRU}{ D. Grumiller and R. McNees, JHEP {\bf 0704} (2007) 074, [arXiv:hep-th/0703230].}
\REF{\CAD}{M. Cadoni and P. Carta, Phys. Lett. {\bf B522} (2001) 126, [arXiv:hep-th/0107234].}
\REF{\CON}{L. Susskind, [arXiv:hep-th/9309145].}
\REF{\SBH}{E. Halyo, A. Rajaraman and L. Susskind, Phys. Lett. {\bf B392} (1997) 319, [arXiv:hep-th/9605112].}
\REF{\EDIH}{E. Halyo, [arXiv:1403.2333]; [arXiv:1406.5763].}
\REF{\EDIO}{E. Halyo, Int. Journ. Mod. Phys. {\bf A14} (1999) 3831, [arXiv:hep-th/9610068]; Mod. Phys. Lett. {\bf A13} (1998), [arXiv:hep-th/9611175].}
\REF{\UNI}{E. Halyo, JHEP {\bf 0112} (2001) 005, [arXiv:hep-th/0108167]; [arXiv:hep-th/0308166].}
\REF{\DES}{E. Halyo, [arXiv:hep-th/0107169]; JHEP {\bf 0112} (2001) 005, [arXiv:hep-th/0108167]; [arXiv:hep-th/0308166].}
\REF{\JT}{R. Jackiw, Nucl. Phys. {\bf B252} (1985) 343; C. Teitelboim, Phys. Lett. {\bf B126} (1983) 41.}
\REF{\SEN}{A. Sen, Entropy {\bf 13} (2011) 1305, [arXiv:1101.4254].}
\REF{\STR}{M. Spradlin and A. Strominger, JHEP {\bf 9911} (1999) 021, [arXiv:hep-th/9904143].}
\REF{\TAK}{T. Azeyanagi, T. Nishioka and T. Takayanagi, Phys.Rev. {\bf D77} (2008) 064005, [arXiv:0710.2956].}
\REF{\REV}{M. Rangamani and T. Takayanagi, Lect.Notes Phys. {\bf 931} (2017) 1, [arXiv:1609.01287].}
\REF{\ENT}{S. Ryu and T. Takayanagi, Phys. Rev. Lett. {\bf 96} (2006) 181602, [arXiv:hep-th/0603001; JHEP {\bf08} (2006) 045,
[arXiv:hep-th/0605073]; T. Nishioka, S. Ryu and T. Takayanagi, JPhys. {\bf A42} (2009) 504008, [arXiv:0905.0932].}
\REF{\VAN}{B. Czech, J. L. Karczmarek, F. Nogueira and M. Van Raamsdonk, Class. Quant. Grav. {\bf 29} (2012) 235025, [arXiv:1206.1323].}
\REF{\MIG}{M. Cadoni and S. Mignemi, Phys. Rev. {\bf D51} (1995) 4319, [arXiv:hep-th/9410041].}
\REF{\RIN}{M. Parikh and P. Samantray, [arXiv:1211.7370].}
\REF{\NHO}{H. K. Kunduri, J. Lucetti and H. S. Reall, Class. Quant. Grav. {\bf 24} (2007) 4169, [arXiv:0705.4214];
P. Figueras, H. K. Kunduri, J. Lucetti and M. Rangamani, Phys. Rev. {\bf D78} (2008) 044042, [arXiv:0803.2998].}
\REF{\DAB}{A. Dabholkar, A. Sen and S. P. Trivedi, JHEP {\bf 0701} (2007) 096, [arXiv:hep-th/0611143].}
\REF{\CQM}{D. Gaiotto, A. Strominger and X. Yin, JHEP {\bf 0511} (2005) 017, [arXiv:hep-th/0412322].}
\REF{\CHI}{A. Strominger, JHEP {\bf 9901} (1999) 007, arXiv:[hep-th/9809027]; V. Balasubramanian, J. de Boer,  M. M. Sheikh--Jabbari and J. Simon, JHEP {\bf 1002} (2010) 017, [arXiv:0906.3272];}
\REF{\SSEN}{R. K. Gupta and A. Sen, JHEP {\bf 0904} (2009) 034, [arXiv:0806.0053]; Entropy {\bf 13} (2011) 1305, [arXiv:1101.4254].}
\REF{\SYK}{A. Kitaev, http://online.kitp.ucsb.edu/online/entangled15/kitaev/, http:// \hfill
online.kitp.ucsb.edu/online/entangled15/kitaev2/,
J. Madacena and D. Stanford, Phys.Rev. {\bf D94} (2016) no.10, 106002, [arXiv:1604.07818]; J. Madacena, D. Stanford and Z. Yang,
PTEP 2016 (2016) no.12, 12C104, [arXiv:1606.01857].}
\REF{\SOL}{S. Solodukhin, Living Rev.Rel. {\bf 14} (2011) 8, [arXiv:1104.3712] and references therein.}
\REF{\REN}{L. Susskind and J. Uglum, Phys. Rev. {\bf D50} (1994) 2700, [arXiv:hep-th/9401070].}
\REF{\JAC}{T. Jacabson, [arXiv:gr-qc/9404039].}
\REF{\CARL}{S. Carlip, Phys. Rev. Lett. {\bf 82} (1999) 2828, [arXiv:hep-th.9812013]; Class. Quant. Grav. {\bf 16} (1999) 3327,
[arXiv:gr-qc/9906126].}
\REF{\SOL}{S. Solodukhin, Phys. Lett. {\bf B454} (1999) 213, [arXiv:hep-th/9812056].}
\REF{\LAST}{E. Halyo, [arXiv:1502.01979]; [arXiv:1503.07808]; [arXiv:1506.05016]; [arXiv:1606.00792].}

\singlespace
\rl{SU-ITP-18/03}
\pagenumber=0
\normalspace
\medskip
\bigskip
\titlestyle{\bf{Schwarzschild Black Holes and Asymptotic $AdS_2$}}
\smallskip
\author{ Edi Halyo{\footnote*{e--mail address: halyo@stanford.edu}}}
\smallskip
\centerline {Department of Physics} 
\centerline{Stanford University} 
\centerline {Stanford, CA 94305}
\smallskip
\vskip 2 cm
\titlestyle{\bf ABSTRACT}
We show that the entropy of D--dimensional \S black holes are given by the entanglement entropy of a boundary of global $AdS_2$ space that lives at asymptotic infinity. We dimensionally reduce General Relativity to two dimensions which leads to
2D dilatonic gravity which has black hole solutions with the same thermodynamics as D--dimensional \S black holes. These dilatonic black holes can be transformed into certain $AdS_2$ black holes by Weyl transformations which are symmetries of the theory that preserve the thermodynamics. In the asymptotic limit, the 
$AdS_2$ black holes become global $AdS_2$ which can be described by two entangled $AdS_2$ Rindler spaces. The entanglement entropy of
a single $AdS_2$ Rindler space reproduces the \S black hole entropy.

\singlespace
\vskip 0.5cm
\endpage
\normalspace

\centerline{\bf 1. Introduction}
\medskip

It is widely believed that the degrees of freedom that describe black hole entropy live near or on the black hole horizon.
This follows from the holographic principle[\HOL] that postulates that the gravitational degrees of freedom that describe the physics in a region live on its boundary. For a black hole this boundary is the horizon. Thus, almost every desription of black hole entropy is basically a description of the horizon or the near horizon region. However, there is another notion of holography in which the degrees of freedom that describe the physics in a region are located on a screen that is far away, e.g. at asymptotic infinity. The celebrated
AdS/CFT correspondence[\ADS] which is the only explicitly holographic theory that exists is an example of this version of holography. An AdS black hole is described by degrees of freedom on the AdS boundary, i.e. the screen and not those on its horizon.

Recently, an attempt was made in ref. [\EDI] to describe \S black hole entropy by degrees of freedom that live at asymptotic infinity. 
It was shown that after a Weyl transformation that preserves the horizon area and an inversion of the radial coordinate, the 
D--dimensional \S metric becomes that of an $AdS_2$ black hole times $S^{D-2}$. At low energies, integrating over $S^{D-2}$ this metric asymptotically becomes global $AdS_2$ which has two boundaries. The $AdS_2$ vacuum corresponds to an entangled state of the two
boundary theories. Concentrating on only one boundary requires tracing over the other one which results in a mixed state with 
nonzero entanglement entropy. The holographic entanglement entropy of a single $AdS_2$ boundary reproduces exactly the entropy of D--dimensional \S black holes. 

Two space-times which are related by a Weyl transformation (in addition to any coordinate transformation) and have the same temperature and entropy have the same thermodynamics. The results of ref. [\EDI] crucially depend on the conjecture that the microscopic entropy counting is the same in such space--times. However, it is well--known that General Relativity(GR) is not symmetric under Weyl transformations[\WALD]. Therefore the space-times generated by a Weyl transformation from the \S metric are not solutions of GR. Thus, the approach of ref. [\EDI] is problematic. Moreover, it is surprising that the entropy of \S black holes which are thermal ensembles is given by the holographic entanglement entropy of global $AdS_2$ which is a zero temperature quantum effect in the vacuum.

In this paper, we relate D--dimensional \S metrics to asymptotically $AdS_2$ spaces in a way that does not suffer from the fact that GR is not Weyl invariant. We first dimensionally reduce D--dimensional GR to two dimensions which results in 2D dilatonic gravity as an effective theory that describes the s--wave sector of GR.
Two dimensional gravity is Weyl (or conformal) invariant since in two dimensions Weyl transformations are just diffeomorphisms. In fact, after imposing the gravity constraints on the 2D metric, the only diffeomorphisms left are the Weyl transformations. The thermal properties such as the temperature and entropy
of two dimensional dilatonic black holes are invariant under Weyl transformations[\GRU]. Thus, 2D dilatonic black holes are divided into conformal classes with the same thermodynamics. In this paper, as in ref. [\EDI], we will assume that 2D black holes in the same class have the same microscopic entropy counting.

The two dimensional dilatonic gravity theory that is obtained from the dimensional reduction of D--dimensional GR has black hole solutions that correspond to the D--dimensional \S black holes with the same temperature and entropy[\CAD]. Using Weyl transformations these can be transformed into certain $AdS_2$ black holes.
In the asymptotic, $r \to \infty$ limit of the original \S coordinates these become global $AdS_2$ spaces.
Then, as in ref. [\EDI], the D--dimensional \S black hole entropy is reproduced by the holographic entanglement entropy of one of the boundaries of global $AdS_2$. 
On the other hand, global $AdS_2$ can also be described as an entangled state of a pair of $AdS_2$ Rindler spaces which is basically the description of $AdS_2$ in an accelarated frame. Each Rindler space is described by one of the boundaries of global $AdS_2$.
An $AdS_2$ Rindler space is just an $AdS_2$ black hole with a radius equal to that of $AdS_2$. Thus, the entanglement entropy of a single, entangled $AdS_2$ Rindler space is simply the entropy of this black hole which reproduces the \S black hole entropy.

This paper is organized as follows. In the next section, we describe the 2D dilatonic gravity, its black hole solutions and their Weyl transformations. In section 3, we transform these black holes into certain
$AdS_2$ black holes. In section 4, we show that, in the asymptotic limit of the original \S cordinates, these spaces reproduce the correct \S black hole entropy either as the holographic entanglement entropy of global $AdS_2$ or the entanglement entropy of a pair of Rindler $AdS_2$ spaces. Section 4 contains a discussion of our results and our conclusions.

\bigskip
\centerline{\bf 2. $D=2$ Dilatonic Gravity and Black Holes}
\medskip

We begin with a brief review of $D=2$ dilatonic gravity, its black hole solutions and Weyl transformations in this theory.
The generic 2D dilatonic gravity action is given by[\GRU]
$$I={1 \over 2} \int d^2x \sqrt{-g} \left[\phi R- U(\phi) (\nabla\phi)^2-2 V(\phi) \right] \quad, \eqno(1)$$
where $\phi$ is the dilaton and $U(\phi)$ and $V(\phi)$ are the kinetic and potential functions respectively. 
This theory has black hole solutions given by the metric and a dilaton profile
$$ds^2=-f(r)dt^2+f(r)^{-1} dr^2 \qquad \phi=\phi(r) \quad, \eqno(2)$$
with
$$f(r)=e^{Q(\phi)} (\omega(\phi)-2M) \qquad {{\partial \phi} \over {\partial r}}= e^{-Q(\phi)} \quad. \eqno(3)$$
The functions $Q(\phi)$ and $\omega(\phi)$ are defined by
$$Q(\phi)=\int^{\phi} d{\bar \phi}~ U({\bar \phi}) \qquad \omega(\phi)=-2\int^{\phi} d{\bar \phi}~V({\bar \phi})e^{Q({\bar \phi})}
\quad. \eqno(4)$$
The black hole horizon is at $\phi_h$ which satisfies
$$\omega(\phi_h)=2M \quad, \eqno(5)$$
and the temperature and entropy of the black hole are given by
$$T_H={{\omega^{\prime}(\phi_h)} \over {4 \pi}} \qquad \qquad S_{BH}=2 \pi \phi_h \quad. \eqno(6)$$
With the normalization of the action in eq. (1), the two dimensional Newton constant is determined by $\phi_h$ to be
$$\phi_h={1 \over {8 \pi G_2}} \quad. \eqno(7)$$
Under Weyl transformations defined by
$$g_{\mu \nu} \to {\hat g}_{\mu \nu}=e^{-2 \sigma(\phi)}g_{\mu \nu} \quad, \eqno(8)$$
the action in eq. (1) becomes
$$I={1 \over 2} \int d^2x \sqrt{-{\hat g}} \left[\phi {\hat R}- {\hat U}(\phi) (\nabla\phi)^2-2 {\hat V}(\phi) \right] \quad, \eqno(9)$$
where the transformed kinetic and potential functions are given by
$${\hat U(\phi)=U(\phi)-2 \sigma^{\prime}(\phi) \qquad {\hat V}(\phi)=e^{2 \sigma(\phi)}} V(\phi) \quad. \eqno(10)$$
As a result, we find that
$${\hat Q}(\phi)=Q(\phi)-2 \sigma(\phi) \qquad {\hat \omega}(\phi)=\omega(\phi) \quad. \eqno(11)$$
Under the Weyl transformation, the form of the black hole solution given by eqs. (2) and (3) remains invariant with $r$ replaced by ${\hat r}$ which is determined by
$\partial_{\hat r}=e^{2{\hat Q}(\phi)} \partial_r$ and $Q$ replaced by ${\hat Q}$. 
For our purposes it is crucial that $\omega(\phi)$ is invariant under Weyl transformations. Then, the black hole 
radius, $\phi_h$ (or equivalently $G_2$) and thus the black hole temperature and entropy are also Weyl invariant. As a result, the thermodynamics of 2D dilatonic black holes is Weyl invariant. Thus, these black holes are divided into equivalency classes 
(with respect to Weyl transformations) with the same thermodynamics.
In the next section, we will use the invariance of 2D dilatonic black hole thermodynamics under Weyl transformations to describe \S black holes in terms of $AdS_2$ black holes.

\bigskip
\centerline{\bf 3. \S Black Holes and Asymptotically $AdS_2$ Spaces}
\medskip

In this section, we relate D--dimensional \S black holes to certain $AdS_2$ black holes. We dimensionally reduce D--dimensional GR 
on $S^{D-2}$ which results in 2D dilatonic gravity which has black hole solutions that correspond to D--dimensional \S black holes. We then Weyl transform these into certain $AdS_2$ black holes with the same temperature and entropy as the original \S black holes. 

We begin by reducing D--dimensional GR on $S^{D-2}$ which describes the s--wave sector of GR in the $t$ and $r$ directions. The dimensional reduction results in 2D dilatonic gravity in which the dilaton parametrizes the volume of the transverse $S^{D-2}$. 

Consider the D--dimensional Einstein--Hilbert action
$$I_{EH}={1 \over 16 \pi G_D} \int d^Dx \sqrt{-g_D}R_D \quad, \eqno(12)$$
where $G_D,g_D$ and $R_D$ are the $D$--dimensional Newton constant, metric and Ricci scalar
respectively. $D$--dimensional Schwarzschild black holes are described by the metrics 
$$ds^2 =-\left(1-{\mu \over r^{D-3}}\right)dt^2+\left(1-{\mu \over r^{D-3}}\right)^{-1}dr^2+r^2d\Omega_{D-2} \quad, \eqno(13)$$   
where 
$$\mu={{16 \pi G_D M} \over {(D-2)A_{D-2}}} \qquad A_{D-2}={{2 \pi^{(D-1)/2}} \over {\Gamma((D-1)/2)}} \quad. \eqno(14)$$ 

We now dimensionally reduce the Einstein--Hilbert action over a $(D-2)$--dimensional sphere of radius $\lambda r=\phi^{-a}$ where
$a=1/(2-D)$ by using the ansatz[\CAD]
$$ds^2=g_{\mu\nu}dx^{\mu}dx^{\nu}+{\phi \lambda^{2-D}}d\Omega^2_{D-2} \quad, \eqno(15)$$
where $\mu,\nu=0,1$ and $x^0=t$, $x^1=r$. Note that the radius over which we dimensionally reduce the theory is not fixed but depends on the dilaton, $\phi$. The constant $\lambda$ is proportional to the inverse Planck length 
$$\lambda=((2(D-2)^{D-2})^{1/(D-3)} \left({{A_{D-2}} \over {16 \pi G_D}} \right)^{1/(D-2)} \quad. \eqno(16)$$
This dimensional reduction results in 2D dilatonic gravity with the action
$$I={1 \over 2} \int d^2x \sqrt{-g} \left[\phi R+ \lambda^2 V(\phi) \right] \quad. \eqno(17)$$
Here $R$ is the two dimensional Ricci scalar and the dilaton potential is given by $V(\phi)=(a+1) \phi^a$. We note that with the normalizaton of the action above, the dilaton is rescaled to be $\phi= A(S^{D-2})/8 \pi G_D$.
The action in eq. (17) has generic black hole solutions given by[\CAD]
$$ds^2=-\left(\phi^{a+1}-{{2M} \over \lambda}\right) dt^2+ \left(\phi^{a+1}-{{2M} \over \lambda}\right)^{-1} dr^2 \eqno(18)$$
and the linear dilaton profile $\phi=\lambda r$. For $a=1/(2-D)$ these correspond to D--dimensional Schwarzschild black holes with horizons at 
$$\phi_h^{a+1}=(\lambda r_h)^{a+1}={{2M} \over \lambda} \quad. \eqno(19)$$ 
The mass, temperature and entropy of these black holes are given by[\CAD]
$$M={\lambda \over 2} \phi_h^{a+1} \qquad \qquad T={\lambda \over {4 \pi}}(a+1) \phi_h^a \qquad \qquad S=2 \pi \phi_h \quad, 
\eqno(20)$$
respectively. By using $\phi_h= A_h/8 \pi G_D$ it is easy to see that
these precisely match the corresponding quantities for D--dimensional Schwarzschild black holes. 

The black hole entropy in eq. (20) can easily be obtained by using the first law of thermodynamics with the Hawking temperature $T_H=f^{\prime}(r)/4 \pi$. One can also, equivalently, obtain the dimensionless Rindler energy, $E_R$, from the near horizon geometry and use the relation $S=2 \pi E_R$[\CON-\DES]. 

We now Weyl transform the theory described by eq. (17) by
$$g_{\mu \nu} \to {\hat g}_{\mu \nu}=(\lambda R)(\phi)^{a-1}g_{\mu \nu} \quad, \eqno(21)$$
so that $\sigma(\phi)=[(1-a)/2] log \phi -(1/2)log(\lambda R)$. $R$ is a free parameter that parametrizes the global scale invariance of the first two terms in the generic 2D action in eq. (9). Of course the dilaton potential, $V(\phi)$, breaks this global scale symmetry so we expect $R$ to appear only in ${\hat V}(\phi)$.
Using eq. (9) we can determine ${\hat U}(\phi)$ and ${\hat V}(\phi)$ and thus the Weyl transformed 
2D dilatonic gravity action
$$I={1 \over 2} \int d^2x \sqrt{-{\hat g}} \left[\phi {\hat R}+ {{(1-a)} \over \phi} \nabla^2 \phi +{{\lambda (a+1)} \over 
{R}} \phi \right] \quad. \eqno(22)$$
Notice that there is now a noncanonical kinetic term for the dilaton and the potential becomes linear just like in the 
Jackiw--Teitelboim theory[\JT] which has $AdS_2$ black hole solutions. 
Computing the  transformed ${\hat Q}(\phi)$ and keeping $\omega(\phi)$ invariant we find 
the new black hole solutions given by 
$$(\lambda R)^{-1}ds^2=-\left(\phi^{2a}-{{2M} \over \lambda} \phi^{a-1}\right) dt^2+ \left(\phi^{2a}-{{2M} \over \lambda} \phi^{a-1}\right)^{-1} 
d{\hat r}^2 \eqno(23)$$
with the dilaton profile $\phi({\hat r})=(a {\hat r}/R)^{1/a}$.
In terms of the new radial coordinate ${\hat r}$ the black hole metric in eq. (23) becomes
$$(\lambda R)^{-1} ds^2=f({\hat r}) dt^2+ f({\hat r})^{-1} dr^2 \qquad, \eqno(24)$$
where
$$f({\hat r})=\left[\left({a{\hat r}} \over R \right)^2-{{2M} \over \lambda} \left({a{\hat r}} \over R \right)^{(a-1)/a} \right] 
\qquad. \eqno(25)$$
Eqs. (24) and (25) describe an $AdS_2$ black hole with a horizon at 
$$\phi_{BH}=\phi_h=\left({{2M} \over \lambda} \right)^{1/(a+1)} \quad. \eqno(26)$$
The $AdS_2$ nature of the metric is fixed by the term $\phi^{a-1}$ in the Weyl transformation in eq. (21) whereas
the $AdS_2$ radius, is determined by the free parameter $R>> \lambda$. Notice that, as expected, $\phi_h$ is the same as in eq. (19) and therefore the entropy (and the temperature) of the black hole in eq. (24) is exactly the same as that in eq. (20) and of the original D--dimensional \S black hole. The new black hole radius is
$${\hat r}_{BH}=\left({{2M} \over \lambda}\right)^{a/(a+1)} {R \over a} \quad. \eqno(27)$$
The physical region for this black hole is given by $0 <{\hat r}< {\hat r}_{BH}$.
This black hole is semiclassical, i.e. ${\hat r}_{BH}>>1/ \lambda$ if the original \S black hole mass is large enough, i.e. 
$2M/\lambda >>(R \lambda/a)^{D-1}$. Since $R$ is a free parameter this can be satisfied for any semiclassical \S black hole with
$M >>1/\lambda$. We also note from eq. (26) that ${\hat r}_{BH} < r_{AdS}=(R/\lambda a)^{1/2}$ in the same range of parameters, so the black hole is smaller than the $AdS_2$ radius. This is
expected since these black holes have negative specific heat and are unstable just like the original \S black holes.

The limit ${\hat r} \to 0$ corresponds to asymprotic infinity in the original Schwarzschild coordinates
since $\phi({\hat r})=(a {\hat r}/R)^{1/a}$ with $a<0$. As ${\hat r} \to 0$, $\phi \to \infty$ and the transverse sphere becomes large signaling that we are approaching asymptotic infinity. Alternatively, we can solve the relation
$\partial_{\hat r}=e^{2Q(\phi)} \partial_r$ to find ${\hat r} \prop r^{1/(2-D)}$ which shows that $r \to \infty$ as ${\hat r} \to 0$.

\bigskip
\centerline{\bf 4. \S Black Hole Entropy from $AdS_2$}
\medskip

We have shown that the $AdS_2$ black holes obtanined in the previous section have the same thermodynamics as that of D--dimensional \S black holes. Our main assumption is that if the thermodynamics of these two space--times are the same, then the microscopic entropy
counting should also be same. 
We now show that the entropy of the $AdS_2$ black holes (and therefore that of the original D--dimensional Schwarzschild black holes) can be obtained either as the holographic
entanglement entropy of a single boundary of global $AdS_2$ or as the entanglement entropy of a Rindler $AdS_2$ space entangled with its thermofield double.

We take the asypmtotic limit ${\hat r} \to 0$ (or ${\hat r}<<r_{BH}$) of the $AdS_2$ black hole metric in eq. (25). In this limit the black hole disappears. In the boundary theory, this correspons to taking the UV limit in which the finite temperature effects that describe the black hole are negligible. Due to the IR/UV duality, this corresponds to the bulk IR limit. 
Then, we are left with the metric
$$ds^2=-{{\hat r}^2 \over r_0^2} dt^2+ {r_0^2 \over {\hat r}^2} dr^2 \quad, \eqno(28)$$
with $r_0=(D-2)\sqrt{R/\lambda}$. This is the metric of the Poincare patch of $AdS_2$. The coordinate transformation[\SEN]
$${\hat r} \pm t=tan{1 \over 2}\left[{1 \over 2}(\sigma \pm \tau) \pm {\pi \over 2} \right] \quad, \eqno(29)$$
takes the Poincare patch into global $AdS_2$ described by the metric
$$ds^2=r_0^2 {{-d \tau^2+d \sigma^2} \over sin^2\sigma} \quad. \eqno(30)$$
In the following, we will use global $AdS_2$ since the Poincare patch has only one boundary and therefore no entanglement entropy. 
Global $AdS_2$, on the other hand, has two disconnected (one dimensional) boundaries at $\sigma=0,\pi$ 
and therefore a nonvanishing entanglement entropy which may reproduce \S black hole entropy. We are allowed to use the transformation in eq.(29) not only classically but also quantum mechanically since the the Poincare $AdS_2$ vacuum seen in global coordinates are
free of particles and vice versa[\STR].

In the $AdS_2$ vacuum, the degrees of freedom on the two boundaries are entangled[\TAK].
The total Hamiltonian of global $AdS_2$ is given by $H_{tot}=H_1-H_2$ where $H_{1,2}$ are the (unknown) Hamiltonians that describe the copies of conformal quantum mechanics living on each boundary. The $AdS_2$ vacuum is a pure but entangled state given by
$$|\Psi_{AdS}>= {1 \over \sqrt{N}}\sum_i |i>_1 \otimes |i>_2 \quad, \eqno(31)$$
where $|i>_1$ ($|i>_2$) is the $N$ vacuum eigenstates of $H_1$ ($H_2$).
If we are restricted to only one boundary, then we need to trace over the states of the second one. As a result, the state in eq. (31) becomes a mixed state described by the density matrix 
$$\rho_1=Tr_2 \rho={1 \over N}Tr_2 \sum_i (|i>_1 \otimes |i>_2) (<i|_2 \otimes <i|_1) \quad.  \eqno(32)$$ 
The entanglement entropy is then given by[\REV]
$$S_{ent}=-Tr(\rho_1log \rho_1)=-{\partial \over \partial n} log (Tr \rho_1^n)|_{n=1} \quad. \eqno(33)$$
Using the holographic entanglement entropy formula, $S_{ent}$ can be computed by[\ENT]. 
$$S_{ent}(A)= {Area(\Sigma_A) \over {4G_2}} \quad, \eqno(34)$$
where $Area(\Sigma_A)$ is the area of the codimension two minimal surface in the bulk, $\Sigma_A$, such that the boundaries of $A$ and $\Sigma_A$ coincide. In our case, $A$ is one of the pointlike boundaries of $AdS_2$ and thus the minimal surface is a point in the bulk with 
$Area(\Sigma_A)=1$.  Therefore, using eq. (34) we get[\TAK]
$$S_{ent}(AdS_2)={1 \over {4 G_2}}=2 \pi \phi_h \quad, \eqno(35)$$
where we used the relation $\phi_h= 1/ {8 \pi G_2}$.
which agrees with eq. (20) and the \S black hole entropy.
The holographic entanglement entropy of global $AdS_2$ can also be computed in more detail by using the methods of refs. [\EDI] which we do not reproduce here. Note that $S_{ent}(AdS_2)$ does not depend on the $AdS_2$ radius which in our case is a free parameter.
It only depends on $G_2$ which is not a constant but determined by the size of the $AdS_2$ black hole in eq. (24) through $\phi_h$. Thus, global $AdS_2$ has a memory of the black hole even though it disappeared in the asymptotic limit. 

We found that \S black hole entropy can be obtained by the entanglement entropy of pure $AdS_2$ which lives at asymptotic infinity.
This entropy is due to the entanglement of the two boundaries of $AdS_2$ and thus purely quantum mechanical. It is surprising that \S black hole entropy which is a thermal effect corresponds to a purely quantum effect in the $AdS_2$ vacuum which is at $T=0$. This is the same result obtained in ref. [\EDI] by different methods. 

However, this result becomes less surprising if we note that global $AdS_2$ can be seen as an entangled state of two copies of
Rindler $AdS_2$. In general, global $AdS_n$ can be described as an entangled state of two zero mass hyperbolic black holes with the
black hole radii equal to the $AdS_n$ radius[\VAN]. When their masses vanish, these hyperbolic black holes simply describe Rindler $AdS_n$ spaces; i.e. $AdS_n$ spaces seen from an accelerated frame with $a=1/r_{AdS}$.  In our case, since the boundary of $AdS_2$ is one dimensional, the hyperbolic nature of the boundary is irrelevant. As a result, global $AdS_2$ is described by two entangled $AdS_2$ Rindler spaces. This is completely analogous to the well--known description of the Minkowski vacuum in terms of two Rindler spaces (or wedges) in a thermofield double state. By holography, $AdS_2$ Rindler spaces are described by their boundary theories. Therefore, the $AdS_2$ vacuum corresponds to the thermofield double
state of the two boundary theories
$$|\Psi_{AdS}>= {1 \over \sqrt{M}}\sum_{i} e^{-\beta E_i/2} |i>_1 \otimes |i>_2 \quad, \eqno(36)$$
where the sum is over all the ($M$) states of the boundary theories since the black hole is an excited (or thermal) state, 
$\beta=2 \pi r_0$ is the inverse Rindler temperature and $E_i$ is the energy of state $|i>$.
Again, if we are restricted to only one boundary, then we need to trace over the states of the second one. As a result, the 
entangled state in eq. (36) becomes a mixed state described by the density matrix 
$$\rho_1={1 \over M} \sum_{i} e^{-\beta E_i}(|i>_1 <i|_1) \quad.  \eqno(37)$$ 
In principle, the entanglement entropy of this mixed state can be computed from the density matrix
in eq. (37) or by using the holographic entanglement entropy prespcription as we did for global $AdS_2$.

In order to compute the entanglement entropy, we will use the fact that $AdS_2$ Rindler spaces are just $AdS_2$ black holes with $r_s=r_{AdS}$. The entanglement entropy  
of the mixed state in eq. (37) is then given by the entropy of the corresponding black hole.

Consider dilatonic $AdS_2$ gravity (i.e. the Jackiw--Teitelboim theory [\JT]) with the action
$$I={1 \over {2}} \int d^2x \phi \left(R+{2 \over L^2}\right) \quad, \eqno(38)$$
where the cosmological constant is given by $\Lambda=-2/L^2$. 
This theory has dilatonic black holes with the metric[\MIG]
$$ds^2= -\left({r^2 \over L^2}- 2 M L \right)dt^2+\left({r^2 \over L^2}- 2 M L \right)^{-1}dr^2
\quad, \eqno(39)$$
and the linear dilaton profile $\phi= r/8 \pi G_2 L$ where again the normalization of the action in eq. (38) has been taken into account. The black hole horizon is at $r_s=(2 M L^3)^{1/2}$. The mass, temperature and entropy of these black holes are given by[\MIG]
$$M_{BH}={r_s^2 \over {2 L^3}} \qquad T_{BH}={r_s \over {2 \pi L^2}} \qquad S_{BH}={r_s \over {4G_2 L}} \quad. \eqno(40)$$

Now, consider an $AdS_2$ black hole with $M=1/2 L$. The metric then becomes 
$$ds^2= -\left({r^2 \over L^2}- 1 \right)dt^2+\left({r^2 \over L^2}-1 \right)^{-1}dr^2  \quad, \eqno(41)$$
which is a black hole with $r_s=L$. This black hole metric also describes global $AdS_2$ given by eq. (30) after the coordinate transformation[\SEN]
$$\tau \pm \sigma=2 tan^{-1} tanh{1 \over 2} \left({t \over L} \pm {1 \over 2}log{{(r/L)-1} \over {(r/L)+1}} \right) \quad. \eqno(42)$$
Eq. (41) actually describes an $AdS_2$ Rindler space with an acceleration $a=1/L$; i.e. the horizon at $r_s=L$ is an accelartion horizon. In order to see this, consider the coordinate transformation
$\rho=\sqrt{r^2-L^2}$ that takes the metric in eq. (41) to[\RIN]
$$ds^2=-{\rho^2 \over L^2}dt^2+\left(1+{\rho^2 \over L^2} \right)^{-1} d\rho^2 \quad. \eqno(43)$$
For $\rho<<L$ the metric describes Rindler space with $a=1/L$ whereas for $\rho>>L$ it becomes that of the Poincare patch of $AdS_2$. 

The entanglement entropy of the two $AdS_2$ Rindler spaces is the entropy of the black hole in eq. (41). From eq. (40) we get 
$$S_{BH}={1 \over {4 G_2}}=2 \pi \phi_h \quad, \eqno(44)$$
where in the second equality we used the relation $\phi_h= 1/8 \pi G_2$.
This is the correct entropy for the two dimensional dilatonic black hole. For a two dimensional theory obtained by dimensional reduction over $S^{D-2}$, $G_2=G_D/A_h$ or
equivalently the horizon value of the dilaton becomes $\phi_h=A_h/8 \pi G_D$. Thus, eq. (44) is exactly the entropy of the original \S black hole.

The description of the \S black hole entropy in terms of two entangled $AdS_2$ Rindler spaces seems more appropriate than the description in terms of the entanglement entropy of global $AdS_2$ since the former is a thermal effect whereas the latter is a pure quantum one. We note that the $AdS_2$ Rindler spaces are at asymptotic infinity in the original \S coordinates and not just inside and outside of the horizon in the conventional entanglement entropy approach to black hole entropy.


\bigskip
\centerline{\bf 5. Conclusions and Discussion}
\medskip

In this paper, we related D--dimensional \S black holes to $AdS_2$ Rindler spaces. Dimensionally reducing GR in D--dimensions 
on an $S^{D-2}$ gives rise to 2D dilatonic gravity which has black hole solutions that correspond to D--dimensional \S black holes. 
We Weyl transformed these dilatonic black holes into $AdS_2$ black holes which, due to the
invariance of 2D black hole thermodynamics under Weyl transformations, have the same entropy (and temperature), In the asymptotic limit, i.e. $r \to \infty$ in the original \S coordinates, these black hole metrics reduce to global $AdS_2$ space which can also be described as two entangled $AdS_2$ Rindler spaces.
$AdS_2$ Rindler space is just an $AdS_2$ black hole with a radius equal to that of $AdS_2$, so the entanglement entropy of a single $AdS_2$ Rindler space is just the 
entropy of this black hole. This matches the entropy of D--dimensional \S black holes. Our results indicate that \S black hole entropy can be located at asymptotic infinity but this can be seen only in a Weyl transformed frame and not the original \S coordinates which is asymptotically flat.

It is well--known that the near horizon geometries of extremal black holes contain an $AdS_2$ factor[\NHO] which is the origin of their entropies[\DAB]. In this paper, we found that \S black holes, with a near horizon geometry that is Rindler space, have an $AdS_2$ factor which appears in their Weyl transformed asymptotic geometries. It would be interesting to see if there is a relation between these two types of black holes due to the fact that $AdS_2$ seems to be the source of their entropies.

In all methods of computing black hole entropy, the degrees of freedom are located on the horizon which is consistent with the holographic principle. However, we showed above that they may also be located at asymptotic infinity. At first thought, this is quite puzzeling.
In order to understand how this can make sense, consider a spherical shell that starts to collapse from infinity and forms a black hole. We assume that, at the beginning of the collapse the shell is entangled with degrees of freedom at infinity and the whole system is in a pure state. After the black hole is formed, we lose access to the collapsing shell's degrees of freedom since they are behind the horizon. As a result, we have to trace over them which leaves us with a mixed state at infinity. In this case, the entropy of the black hole is completely due to the entanglement of the shell which is behind the horizon and
the degrees of freedom at infinity. Thus, the entanglement entropy of the degrees of freedom at infinity is the entropy of the black hole. Our results constitute suporting evidence for this idea. 
 
Even though we found that the entropy of \S black holes is given by the entanglement entropy of two $AdS_2$ Rindler spaces, we do not have a clear idea about the degrees of freedom that we count. 
Following the AdS/CFT correspondence, we expect that the boundary of $AdS_2$ is described by a one dimensional CFT with only a time coordinate, i.e. conformal quantum mechanics[\CQM] which is not well--understood. However, by compactifying the much better understood $AdS_3$ to $AdS_2$ (on its boundary $S^1$), this theory was shown to be equivalent to a chiral or light--cone 2D CFT[\CHI]. It seems that one can also count the $AdS_2$ entropy in certain situations in string theory[\SSEN]. The most promising description of the (near) $AdS_2$ boundary theory seems to be the SYK model[\SYK]. It has many of the properties that belong to black holes such as chaos with a maximal Lyapunov exponent. Needless to say, the nature of the one--dimensional boundary theory dual to $AdS_2$ is a very important subject that requires further investigation. 

It is interesting to compare our results with the more conventional derivation of \S black hole entropy as entanglement entropy which has a rich literature[\SOL]. First, the conventional entanglement entropy describes the entanglement between degrees of freedom just inside and outside the horizon. In our case, the entangled degrees of freedom belong to two $AdS_2$ Rindler spaces that live at asymptotic infinity. These are similar to the two Rindler wedges that describe the near horizon region of an analytically extended black hole Penrose diagram but with different
asymptotics. Unlike the \S black hole case, the $AdS_2$ Rindler space has a boundary and is therefore holographic.
Second, the conventional entanglement entropy is a UV cutoff dependent quantity that diverges in the continuum limit. 
It is a one--loop correction to the black hole entropy[\REN] and reproduces it completely only in the context of induced gravity[\JAC].
In our case, the entanglement entropy of two $AdS_2$ Rindler spaces is a cutoff independent finite quantity that gives the correct result. It seems that $AdS_2$ naturally introduces a finite (and correct) cutoff. 
Third, conventional entanglement entropy depends on the type and number of quantum fields that are assumed to live near the horizon which leads to the species problem. In our case, we do not need to know the degrees of freedom on the $AdS_2$ boundary to compute the entanglement entropy of the two $AdS_2$ Rindler spaces since it is given by that of a specific $AdS_2$ black hole. Clearly, there is no species problem.


It has been shown that the near horizon region with the geometry of Rindler space can be described by a CFT that  reproduces \S black hole entropy[\CARL,\SOL,\LAST]. The results of this paper provide an alternative description of black hole entropy in terms of degrees of freedom that live at infinity. These two descriptions differ by the replacement of Rindler space by $AdS_2$ Rindler space. Both of these theories are described by 2D CFTs (even though $AdS_2$ Rindler space is holographic and has a dual description in terms of the one dimensional boundary theory). Rindler space is related to a $AdS_2$ Rindler space by a coordinate and Weyl transformation[\SEN]. 
It is tempting to think that the black hole is described by both of these theories which are connected by a renormalization flow from infinity to the horizon.

\endpage

\bigskip
\centerline{\bf Acknowledgments}

I would like to thank the Stanford Institute for Theoretical Physics for hospitality.
I would also like to thank Lenny Susskind and Shamit Kachru whose questions provided the motivation for this paper.

\vfill

\refout

\end
\bye